\documentclass[letterpaper,11pt]{article}
\pdfoutput=1 % if your are submitting a pdflatex (i.e. if you have
\usepackage{jcappub} % for details on the use of the package, please
\usepackage[T1]{fontenc} % if needed

\def\beq{\begin{equation}}
\def\eeq{\end{equation}}

\title{Bayesian Analysis of a Generalized Starobinsky Model with Reheating Constraints}

\author[a]{Francisco X. Linares Cede\~no}
\author[b]{Gabriel Germ\'an,}
\author[b]{Juan Carlos Hidalgo}

\affiliation[a]{Instituto de F\'isica y Matem\'aticas, Universidad Michoacana de San Nicol\'as de Hidalgo,\\ Edificio C--3, Ciudad Universitaria, CP. 58040 Morelia, Michoac\'an, M\'exico,}
\affiliation[b]{Instituto de Ciencias F\'{\i}sicas, Universidad Nacional
Aut\'onoma de M\'exico,\\ Av. Universidad s/n, Cuernavaca, Morelos, 62210, M\'exico.}

\emailAdd{francisco.linares@umich.mx}
\emailAdd{gageve@gmail.com}
\emailAdd{hidalgo@icf.unam.mx}

\abstract{We study a generalization of the the Starobinsky model adding a term of the form  $R^{2p}$ to the Einstien-Hilbert action. We take the power $p$ as a parameter of the model and explore the constraints from CMB plus BAO data through a Bayesian analysis, thus exploring a range of values for the exponent parameter. We incorporate a reheating phase to the model through the background matter content (equation of state) and the duration of this period (number of $e$-foldings of reheating). We find that incorporating information from reheating imposes constraints on cosmological quantities, more stringent than the case of no reheating when tested with the Planck+BAO data. The inferred value of the exponent parameter is statistically consistent with $p=1$, favoring the original Starobinsky potential. Moreover, we report tighter constraints on $p$ and the number of $e$-folds in comparison with previous works. The obtained values for other inflationary observational parameters, such as the scalar spectral index $n_s$ and the scalar amplitude of perturbations $A_s$, are consistent with prior measurements. Finally we present the alternative use of consistency relations in order to simplify the parameter space and test the generalized Starobinsky potential even more efficiently.}

\begin{document}
\maketitle
\flushbottom

%%%%%%%%%%%%%%%%%%%%%%%%%%%%%%%%%%%%%%%%%%%%
%%%%%%%%%%%%%%%%%%%%%%%%%%%%%%%%%%%%%%%%%%%%
\section{Introduction}\label{Intro}
While the origin of the Universe is largely unknown, a lot of understanding has been gained through the hypothesis known as the "inflationary paradigm" \cite{Linde:1984ir,Lyth:1998xn,Baumann:2009ds,Martin:2018ycu,Odintsov:2023weg}. According to this hypothesis, our Universe experienced a rapid expansion at its inception, followed by a reheating period that led to the hot Big Bang (reviews on reheating are found e.g., un Refs.~\cite{Bassett:2005xm,Allahverdi:2010xz,Amin:2014eta}). Inflation is typically described by a scalar field referred to as the "inflaton" field. After approximately 380,000 years, the early universe became transparent to photons which then produced what we observe today as the Cosmic Microwave Background radiation (CMB). This probe of the early universe provides valuable information regarding the physics of the inflationary era \cite{Planck:2018jri}. In particular, the inflationary paradigm predicts the presence of primordial gravitational waves; tensor fluctuations in spacetime \cite{LISACosmologyWorkingGroup:2023njw}. Numerous experiments have been conducted to detect such waves, and have managed to impose upper limits on their amplitude \cite{CMB-S4:2020lpa}. Understanding the dynamics of the inflaton field and its realizations is crucial for interpretation the observed CMB data.

Among the variety of inflationary models lies the scalar field potential belonging to the Starobinsky type. The Starobinsky model~\cite{Starobinsky:1980te} emerged as one of the early proposals for inflation. It is characterized by an action that includes the Einstein-Hilbert term, together with a scalar curvature term of higher order. The action takes the form
\begin{equation}
\label{Staropot}
S=\frac{M_{Pl}^{2}}{2}\int d^{4}x \sqrt{-g}\left(R+\frac{1}{6M^2} R^{2}\right),
\end{equation}
In this equation, $R$ denotes the Ricci scalar, $M_{Pl}=2.44\times 10^{18} \,\mathrm{GeV}$ is the reduced Planck mass, and $M$ corresponds to the Starobinsky free parameter, expressed in terms of mass units. Similar to any $f(R)$ theory, the Starobinsky action can be transformed into a form that includes the Einstein-Hilbert term and an action for a scalar field. This is achieved by considering a conformal transformation of the metric
\begin{equation}
g_{\mu\nu}\rightarrow e^{\sqrt{\frac{2}{3}}\frac{\phi}{M_{Pl}}}g_{\mu\nu}.
\end{equation}
Upon this transformation, we obtain the following action for the Starobinsky model:
\begin{equation}
S=\int d^{4}x \sqrt{-g}\left(\frac{M_{Pl}^{2}}{2}R-\frac{1}{2}\partial_{\mu}\phi\partial^{\mu}\phi-V(\phi)\right),
\end{equation}
where $V(\phi)$ represents the potential of the scalar field, given by
\begin{equation}
\label{Spot}
V(\phi)=V_{0}\left(1-e^{-\sqrt{\frac{2}{3}}\frac{\phi}{M_{Pl}}}\right)^{2}.
\end{equation}
Here $V_{0}$ coreresponds to the product $\frac{3}{4}M_{Pl}^{2}M^2$. 

In this paper we use Bayesian analysis to explore a generalization of this model. Specifically the generalized Starobinsky model which adds a term of the from $R^{2p}$  to the Einstein-Hilbert action \cite{Motohashi:2014tra}. The study includes an examination of a range of values for the exponent, as part of the parameter space. Our analysis presents a comprehensive exploration of the parameter space, with a particular emphasis on constraining  { the potential amplitudes and curvatu\rm re} through numerical evolution of the Boltzmann equation. The statistical analysis yields posterior distributions for the model parameters, together with cosmological parameters such as the number of $e$--folds, spectral index, and the tensor-to-scalar ratio. 

The article is organized as follows: Section~\ref{GStaro} provides an overview of the  $R^{2p}$  generalized Starobinsky model. Section~\ref{Bayes} presents the results of the statistical analysis using Bayes' theorem, including the posterior distributions and the Bayes' factor for model selection. We provide a discussion with final remarks in Section~\ref{Con}.

%%%%%%%%%%%%%%%%%%%%%%%%%%%%%%%%%%%%%%%%%%%%
%%%%%%%%%%%%%%%%%%%%%%%%%%%%%%%%%%%%%%%%%%%%
\section{\boldmath $R^{2p}$ Generalization of the Starobinsky model}\label{GStaro}
%%%%%%%%%%%%%%%%%%%%%%%%%%%%%%%%%%%%%%%%%%%%
%%%%%%%%%%%%%%%%%%%%%%%%%%%%%%%%%%%%%%%%%%%%

\begin{figure}[ht!]
\centering
  \includegraphics[width=.82\linewidth]{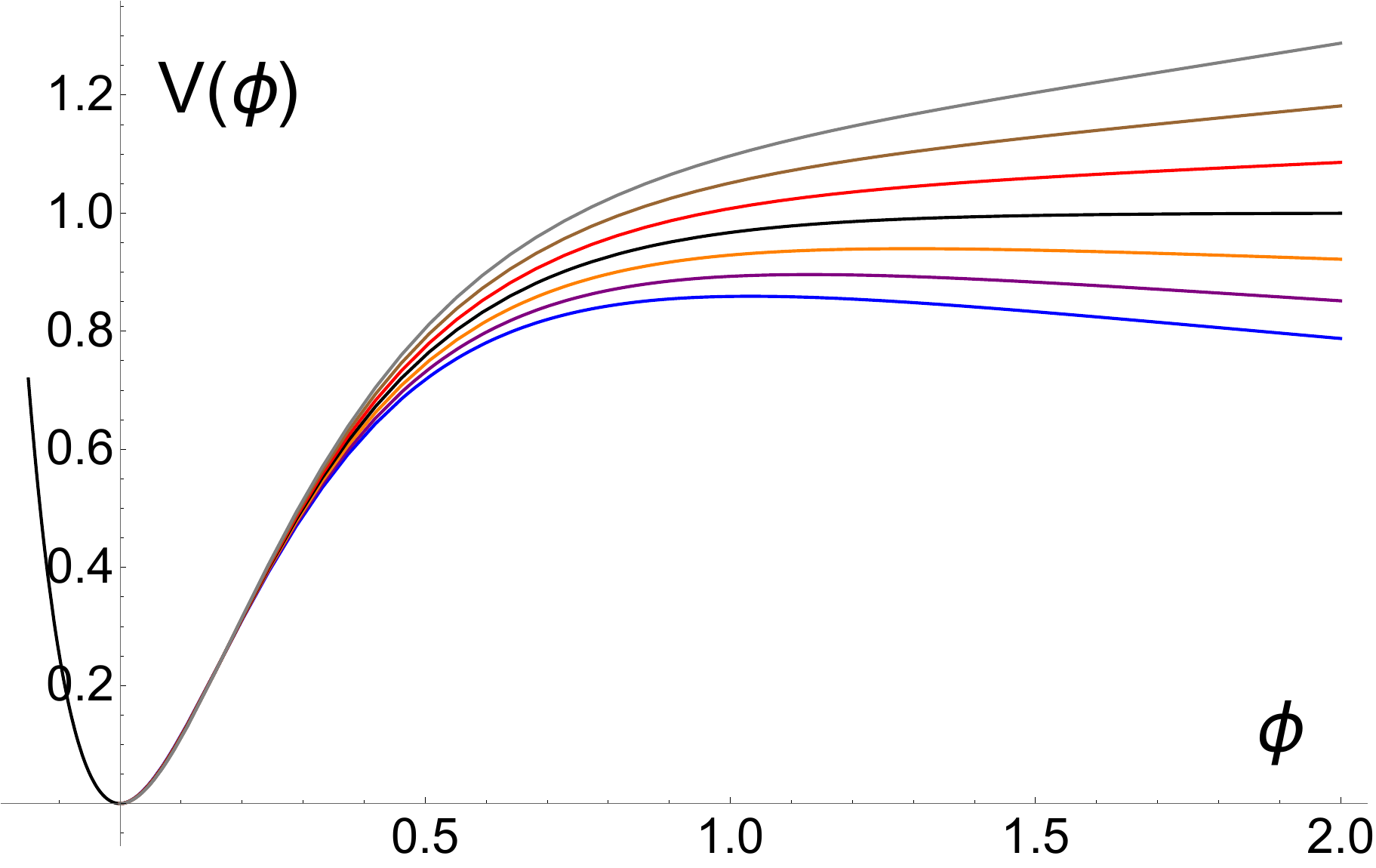}
  \caption{The figure illustrates the graph of the generalized Starobinsky potential, scaled in units of $V_0$, as defined in Eq.~(\ref{genpot}). The plot illustrates the potential for various values of $p$, namely (from top to bottom): $p=0.985, 0.99, 0.995, 1, 1.005, 1.01, 1.015$.}
  \label{pot}
\end{figure}

The generalized model in question stems from a straightforward extension of the Starobinsky action
\begin{equation}
S_{\text{Gen}}=\frac{M_{Pl}^{2}}{2}\int d^{4}x \sqrt{-g}\left(R+\left(6M^2\right)^{\frac{1}{1-2p}} R^{2p}\right),
\label{sgen}
\end{equation}
which recovers the original Starobinsky model when $p=1$.
In the Einstein Frame, the model gives rise to a generalised potential as follows
\begin{equation}
V= V_0e^{-2\sqrt{\frac{2}{3}}\frac{\phi}{M_{Pl}}}\left(e^{\sqrt{\frac{2}{3}}\frac{\phi}{M_{Pl}}}-1\right)^{\frac{2p}{2p-1}},
\label{genpot}
\end{equation}
where $V_{0}= \frac{2p-1}{2}\left(\frac{1}{2p}\right)^{\frac{2p}{2p-1}}\left(6M^2\right)^{\frac{1}{2p-1}}M_{Pl}^2$ (see Fig.~\ref{pot}).
%%%%%%%%%%%%%%%
%%%%%%%%%%%%%%%
%%%%%%%%%%%%%%%
%%%%%%%%%%%%%%%

This generalization of the Starobinsky model has been studied in various contexts. For example, Motohashi et al.  \cite{Motohashi:2014tra}  explored the consistency relations of the generalized model at first order, while Renzi et al. \cite{Renzi:2019ewp} investigated the stability of predictions for $r$, taking into account experimental uncertainties on $n_s$ and assuming the validity of $\Lambda$CDM.
Here we follow a similar approach to that of Ref.~\cite{Garcia:2023tkk}. 

Cosmological observables in the context of inflationary models can be described through expressions derived from the slow-roll (SR) approximation, as demonstrated in previous studies (see, for example, \cite{Lyth:1998xn}, \cite{Liddle:1994dx}). These are given by
\begin{eqnarray}
n_{t} &=&-2\epsilon = -\frac{r}{8} , \label{Int} \\
n_{s} &=&1+2\eta -6\epsilon ,  \label{Ins} \\
\alpha &=&16\epsilon \eta -24\epsilon^2 - 2\xi_2 ,  \label{alpha} \\
A_s(k) &=&\frac{1}{24\pi ^{2}} \frac{V}{\epsilon\, M_{Pl}^{4}}. \label{IA} 
\end{eqnarray}

In these equations, $n_t$ represents the tensor spectral index, $r$ is the tensor-to-scalar ratio, $n_s$ corresponds to the scalar spectral index and $\alpha$ its running. The amplitude of density perturbations at a specific wavenumber $k$ is denoted by $A_s(k)$. These quantities are evaluated at the time when the wavenumber $k$ crosses the horizon. The SR parameters introduced above are defined as
\beq
\epsilon \equiv \frac{M_{Pl}^{2}}{2}\left( \frac{V^{\prime }}{V }\right) ^{2},\quad\quad
\eta \equiv M_{Pl}^{2}\frac{V^{\prime \prime }}{V}, \quad\quad \xi_2 \equiv M_{Pl}^{4}\frac{V^{\prime }V^{\prime \prime \prime}}{V^2}.
\label{Spa}
\eeq
Of course, primes on $V$ indicate derivatives with respect to the inflaton field $\phi$.
%%%%%%%%%%%%%%%%%%%%%%%%%%%%%%%%%%%%%%%%%%%%
\subsection{Consistency relations}
%%%%%%%%%%%%%%%%%%%%%%%%%%%%%%%%%%%%%%%%%%%%
The number of $e$--folds during inflation, $N_{k} \equiv \ln\left(\frac{a_{e}}{a_{k}}\right)= -\frac{1}{M_{Pl}^2}\int_{\phi_k}^{\phi_e}\frac{V}{V'}d\phi$, is explicitly
\beq
\label{genNk}
N_k=\frac{\sqrt{6}}{4}\left(\frac{\phi_e}{M_{Pl}}-\frac{\phi_k}{M_{Pl}}\right)+\frac{3p}{4\left(p-1\right)}\ln \left(\frac{2p-1-\left(p-1\right)e^{\sqrt{\frac{2}{3}}\frac{\phi_e}{M_{Pl}}}}{2p-1-\left(p-1\right)e^{\sqrt{\frac{2}{3}}\frac{\phi_k}{M_{Pl}}}}\right),
\eeq
where $\phi_{e}$ is given by the solution of the condition $\epsilon=1$
\beq
\phi_{e}=\sqrt{\frac{3}{2}} M_{Pl} \ln \left[\frac{1-4 p^2-2 \sqrt{3} p (2 p-1)}{1+4 p-8 p^2}\right],
\eeq
and $\phi_k$ is obtained by solving e.g., $16\epsilon=r$. This evaluation yields
\beq
\label{genfik}
 \phi_{k}=\sqrt{\frac{3}{2}} M_{Pl} \ln \frac{\left(2p-1\right)\left[64-3r+p\left(-64+8\sqrt{3r}+6r\right)\right]}{3\left(2p-1\right)^2r-64\left(p-1\right)^2}.
\eeq
With this expression for the inflaton and, using the equation for the spectral index Eq.~(\ref{Ins}), we find $p$ in terms of $r$ and $n_s$
\beq
\label{p}
p=\frac{\left(8+\sqrt{3r}\right)^2}{3\left(4+\sqrt{3r}\right)^2-8\left(1-3n_s)\right)}.
\eeq
In our case, it is more convenient to express this relation for $r$ in terms of $n_s$ and $p$ as follows
\beq
\label{genr}
r=\frac{8\left(8+2\sqrt{2}(2-3p)\sqrt{p(5+n_s(3-9p)+3p)}+p(-19+n_s(3-9p)+21p)\right)}{3\left(3p-1\right)^2}.
\eeq
Using the equation for $\alpha$ we have
\beq
\label{genalpha}
\alpha=\frac{\sqrt{r}(r - 8+8n_s)(14-6n_s+3\sqrt{3r})}{16(8\sqrt{3}+3\sqrt{r})}\, .
\eeq
The overall scale of the potential, $V_0$, is obtained by solving the equation for the amplitude of scalar perturbations (\ref{IA}) at horizon crossing. The result is
\begin{equation}
V_0= \frac{3}{2}A_s\,\pi^2\,r\,e^{2\sqrt{\frac{2}{3}}\frac{\phi_k}{M_{Pl}}}\left(e^{\sqrt{\frac{2}{3}}\frac{\phi_k}{M_{Pl}}}-1\right)^{-\frac{2p}{2p-1}}.
\label{V0}
\end{equation}

%%%%%%%%%%%%%%%%%%%%%%%%%%%%%%%%%%%%%%%%%%%%
\subsection{Reheating bounds on cosmological parameters}
%%%%%%%%%%%%%%%%%%%%%%%%%%%%%%%%%%%%%%%%%%%%

The quantity representing the number of $e$--folds during reheating, commonly referred to as $N_{\rm re}$, is typically described using the logarithm of scale factors from the end of inflation to the end of reheating. Alternatively, it can be defined in relation to the energy densities at the end of inflation and at thermalization
\begin{equation}
N_{\rm re}= \ln\left(\frac{a_{\rm re}}{a_{e}}\right)=\frac{1}{3(1+\omega_{\rm re})} \ln\left(\frac{\rho_e}{\rho_{\rm re}}\right).
\label{Nre1}
\end{equation}
The second equality can be derived from the fluid equation's solution, $\rho\propto a^{-3(1+\omega_{\rm re})}$, assuming a constant effective equation of state (EoS) parameter $\omega_{\rm re}$. Using the first equation in (\ref{Nre1}), we find  \cite{German:2023yer}
\begin{equation}
N_{\rm re}=\ln\left(\frac{\left(\frac{43}{11g_{s,\rm re}}\right)^{1/3}\sqrt{A_s\,r}\,\pi M_{Pl}T_0\,e^{-N_k}}{\sqrt{2}\,k_pT_{\rm re}}\right).
\label{Nre2}
\end{equation}
Furthermore, we can express the corresponding energy densities as follows
\begin{equation}
\rho_{e}=\frac{3}{2}V_e, \quad\quad \rho_{\rm re}=\frac{\pi^2 g_{\rm re}}{30}T_{\rm re}^4,
\label{ros}
\end{equation}
Here, $V_e$ represents the inflationary potential at the end of inflation. Without loss of generality, we can express the potential as $V(\phi)=V_0 f(\phi)$, where $V_0$ is the overall scale and $f(\phi)$ contains the factor dependent on $\phi$. By writing $V_e=\frac{V_e}{V_k}V_k= 3\frac{V_e}{V_k}H_k^2M_{Pl}^2$, we obtain
\begin{equation}
V_e=\frac{3}{2}\pi^2A_s\, r \frac{f(\phi_e)}{f(\phi_k)} M_{Pl}^4.
\label{Ve}
\end{equation}

In this context, $\phi_k$ (and $\phi_e$) represents the value of the inflaton at horizon crossing (at the end of inflation), while $A_s$ denotes the amplitude of scalar perturbations. By substituting equations (\ref{ros}) and (\ref{Ve}) into equation (\ref{Nre1}), we can derive an alternative expression for $N_{\rm re}$
\begin{equation}
N_{\rm re}=\frac{1}{3(1+\omega_{\rm re})} \ln\left(\frac{135 A_s\ r M_{Pl}^4 f(\phi_e)} {2 g_{\rm re} T_{\rm re}^4 f(\phi_k)}\right).
\label{Nre4}
\end{equation}
By combining equations (\ref{Nre2}) and (\ref{Nre4}), we can derive an expression for $\omega_{\rm re}$ in terms of $n_s$
\begin{equation}
\omega_{\rm re}=-1-\frac{1}{3}\frac{\ln\left(\frac{135A_sM_{Pl}^4 f(\phi_e)r}{2g_{\rm re}T_{\rm re}^4f(\phi_k)}\right)}{N_k-\ln\left(\frac{\left(\frac{43}{11g_{s,\rm re}}\right)^{1/3}\pi\sqrt{A_s}M_{Pl}T_0\sqrt{r}}{\sqrt{2}k_pT_{\rm re}}\right)},
\label{wns}
\end{equation}
In this expression, the dependence on the spectral index $n_s$ arises from the consistency relation (\ref{genr}), which relates $r$ to $n_s$ and $p$.

We examine the constraints provided by Table 3 of  \cite{Planck:2018jri}  for the $\Lambda$CDM+$r+(dn_s/d\ln k)$ cosmological model. The constraints are derived from the Planck TT, TE, EE+lowE+lensing+BK15+BAO dataset, providing limits on the parameters and observables within this specific cosmological model and data combination. The bounds are as follows
\begin{equation}
n_s =0.9658\pm 0.0040\quad (68\%,\mathrm{C.L.}),
\label{boundsns}
\end{equation}
\begin{equation}
r < 0.068\quad (95\%,\mathrm{C.L.}).
\label{boundsr}
\end{equation}
At $2\sigma$, the bounds on $p$ are given by
\begin{equation}
0.990 < p < 1.005.
\end{equation}

The crucial parameter for imposing theoretical constraints in reheating is the equation of state parameter, $\omega_{\rm re}$, as given by Eq.(\ref{wns}), while the remaining quantities displayed in Table~\ref{tabla} are derived by imposing the condition $0<w<0.25$. We immediately note that the spectral index $n_s$ remains unconstrained within the $2\sigma$ range $0.9578 < n_s < 0.9738$, while $p$ or, equivalently, $r$ (through the consistency relation (\ref{p})), is restricted. From there, constraints for the running $\alpha$ and other quantities follow. 

Let us now discuss the choice of range for $\omega_{\rm re}$.
After the inflationary phase, the equation of state parameter (EoS) averaged over the oscillations of the inflation field is approximately $\omega_{\rm re}=0$. However, a significant transition occurs at the end of the preheating phase \cite{Dufaux:2006ee}. Within a brief time interval the EoS undergoes a sharp change from $\omega_{\rm re}=0$ to a value approximately ranging from $\omega_{\rm re}\approx 0.2-0.3$ \cite{Podolsky:2005bw}  (see also \cite{Torres-Lomas:2014bua}). 
The equation of state (EoS) does not immediately reach the radiation-dominated value of 1/3; instead, it settles around $\omega_{\rm re}\approx 1/4$, as depicted in figure 1 of \cite{Podolsky:2005bw}, and maintains that value until complete thermalization occurs.

The departure from an immediate radiation-dominated regime after preheating can be attributed to two facts. Firstly, the light field retains a considerable induced effective mass due to interactions. Secondly, there is a substantial residual contribution from the homogeneous inflaton field. 
These factors contribute to the deviation from an immediate transition to radiation dominance. Therefore, it is reasonable to consider the range $0 < \omega_{\rm re} < 0.25$ for the equation of state during reheating. In the original $2\sigma$ Planck bounds for the spectral index $n_s$ remain unaffected. However, the constraints for all other quantities become more stringent, as outlined in Table \ref{tabla}. In the case of the generalized Starobinsky model, there is no inherent assumption that the reheating temperature  would have the same value as in the original Starobinsky model. Therefore, it is crucial to investigate how this temperature varies when considering a generalized potential. This is detailed in the Appendix of \cite{Garcia:2023tkk}, and reveals that within a narrow range of values around $p=1$, it is still appropriate to use $T_{\rm re}=3.1\times 10^9\,\mathrm{GeV}$ \cite{Gorbunov:2010bn} when estimating the constraints provided in Table \ref{tabla}.

%%%%%%%%%%%%%%%
%%%%%%%%%%%%%%%
\begin{table*}[h!]\label{reheating bounds}
\begin{center}{
\begin{tabular}{|c|c|c|} 
 \hline 
Quantity & Theoretical bounds from reheating & Defining equation \\ \hline 
$\omega_{\rm re}$ & $0 < \omega_{\rm re} < 0.25 $ & Eq.~(\ref{wns}) \\ 
$p$   &   $0.990< p < 1.005$ &Eq.~(\ref{p})  \\ 
$V_0$   &   $3.7\times 10^{-13} > V_0> 1.3\times 10^{-13}$ &Eq.~(\ref{V0})  \\ 
$n_s$   &     $0.9738 > n_{s} >0.9578 $ &Eq.~(\ref{Ins}) \\
$\alpha$   &     $-6.9\times 10^{-4} > \alpha >-6.2\times 10^{-4} $ &Eq.~(\ref{genalpha}) \\ 
$r$   &   $0.0080> r > 0.0024$ &Eq.~(\ref{Int})\\ 
$N_{\rm re}$   &     $18.4> N_{\rm re} > 14.5 $ &Eq.~(\ref{Nre4}) \\
$N_{k}$   &     $51.2 < N_{k} < 54.6 $ &Eq.~(\ref{genNk}) \\
\hline 
 \end{tabular} \\ 
 }
\caption{\label{tabla} Theoretical constraints on the observables, namely $n_s$, $r$, $\alpha$ as well as the number of $e$--folds during reheating $N_{\rm re}$ and during inflation $N_k$, are examined for the generalized Starobinsky model. The initial bounds for $n_s$ and $r$ are provided by Eqs.~(\ref{boundsns}) and (\ref{boundsr}),~\cite{Planck:2018jri}. At the $2\sigma$ level, the range for $n_s$ is presented in the table and is not influenced by the reheating condition. However, the additional constraints stem from enforcing the range $0 < \omega_{\rm re} < 0.25$ on the equation of state during reheating.}
\end{center}
\end{table*}
%%%%%%%%%%%%%%%
%%%%%%%%%%%%%%%

\section{Bayesian analysis}\label{Bayes}

In this Section we perform the corresponding statistical analysis of the generalized Starobinsky potential~\eqref{genpot}. Whereas the main parameters to be constrained with MCMC method are $\log V_0\, , p\, , N_k\, ,$ and $ n_s$, we will use the consistency relations (\textit{cr}) from Eqs.~\eqref{genr} and~\eqref{genalpha} to obtain the posterior probability distribution for the tensor--to--scalar ratio $r^{cr}$ and the scalar running $\alpha^{cr}$, respectively. For that matter, we have used the Boltzmann code \textsc{class}~\cite{Lesgourgues:2011re}, implementing the generalized Starobinsky potential the \texttt{primordial} module, and the software \textsc{Monte Python}~\cite{Audren:2012wb} to perform the Bayesian inference.

Particularly we will be interested in comparing the theoretical bounds from reheating with the observational constraints from the latest Planck Satellite Collaboration data~\cite{Planck:2019nip}, including the likelihoods for temperature ($TT$) power spectra, the polarisation spectra ($TE$ and $EE$), and CMB lensing measurements. Besides, we combine such Planck data with Baryon Acoustic Oscillation (BAO) observations\footnote{For reference, qe will refer to the combination of all these sets of experiments as Planck+BAO.}: the final galaxy clustering data set of the Baryon Oscillation Spectroscopic Survey~\cite{BOSS:2016wmc}, the BAO signal from large-scale correlation function of the 6dF Galaxy Survey (6dFGS)~\cite{Beutler:2011hx}, and a sample of galaxies with low redshift ($z < 0.2$) from the Sloan Digital Sky Survey (SDSS) Data Release 7 (DR7)~\cite{Ross:2014qpa}. 

Table~\ref{stats_priors} shows the priors on the parameters.  The MCMC was performed for those parameters dubbed with role ``\textit{Cosmological}'' (see last column), whereas the scalar spectral index $n_s$ is obtained as a ``\textit{Derived}'' parameter from the previous ones. On the other hand, the tensor--to--scalar ratio $r^{cr}$ and the scalar running $\alpha^{cr}$ are obtained from the consistency relations~\eqref{genr} and~\eqref{genalpha} respectively. 
\begin{table*}[h!]
\begin{center}{
\begin{tabular}{|c|c|c|c|c|} 
 \hline 
Param & mean & min & max & role \\ \hline 
$\log V_{0 }$ & $-22$ & $-34$ & $-10$ & Cosmological \\ 
$p$ &$1$ & $0.9$ & $1.05$ & Cosmological \\ 
$N_k$ &$60$ & $20$ & $100$ & Cosmological \\ 
$n_{s }$ & $1$ & None & None & Derived \\
$r^{cr}$ &$0$ & None & None & Consistency relation~\eqref{genr} \\ 
$\alpha^{cr}$ &$0$ & None & None & Consistency relation~\eqref{genalpha} \\
\hline 
 \end{tabular} \\ 
 }
\caption{\label{stats_priors} Set of priors for the cosmological parameters. From left to right, each column indicates the parameter of interest (Param), the initial value to start the exploration of the parameter space (mean), the minimum and maximum values of the range given to the prior (min and max respectively), and the role of the parameter (role). See text for more details.}
\end{center}
\end{table*}

In order to compare with results from previous analysis in the literature, we use the same priors on $p$ and $N_k$ as those used in~\cite{Renzi:2019ewp}. For the rest of the standard cosmological parameters ($\Omega_b h^2\, ,\Omega_{cdm} h^2\, ,100\theta_s\, ,\tau_{reio}$) we have considered fixed values from Planck 2018\footnote{From~\cite{Planck:2018vyg} we take $\Omega_b h^2 = 0.0224\, ,\Omega_{cdm} h^2 = 0.120\, ,100\theta_s = 1.0411\, ,\tau_{reio} = 0.054\, .$}, since such parameters do not change considerably from the observed values even when the priors are not specified. That is, the dynamics of the inflaton with the generalized Starobinsky potential does not alter the observed values for the matter content (baryonic $\Omega_b h^2$ and cold dark matter $\Omega_{cdm} h^2$), angular diameter distance $100\theta_s$, or the reionization optical depth $\tau_{reio}$.

The statistical constraints on the parameters are shown in Table~\ref{stats_results}. The only parameter with no restrictions within the broad prior we have considered is (the logarithm of) the amplitude of the inflationary potential (first row). For the rest of the parameters we found their corresponding mean values and confidence levels (CL). It is important to notice that, while the consistency relation~\eqref{genalpha} does not show explicit dependence on $p$, by expressing $r$ as function of $p$ and $n_s$ in equation~\eqref{p}, and substituting  in equation~\eqref{genalpha}, we obtain the function $\alpha^{cr} = \alpha^{cr}(p, n_s)$.  
\begin{table*}[h!]
\begin{center}{
\begin{tabular}{|c|c|c|c|c|} 
 \hline 
Param & best-fit & mean$\ \pm\ \sigma$ & 95\% lower & 95\% upper \\ \hline 
$\log V_{0 }$ & $--$ & $--$ & $--$ & $--$ \\ 
$p$ &$1.002$ & $1.003_{-0.0014}^{+0.0011}$ & $1$ & $1.005$ \\ 
$N_k$ &$60.25$ & $59.77_{-2.6}^{+2.6}$ & $54.53$ & $64.97$ \\ 
$n_{s }$ &$0.9645$ & $0.9639_{-0.0027}^{+0.0035}$ & $0.9576$ & $0.9699$ \\ 
$r^{cr}$ &$0.002405$ & $0.002387_{-3.7\times 10^{-5}}^{+4.1\times 10^{-5}}$ & $0.002308$ & $0.002459$ \\ 
$\alpha^{cr}$ &$-5.216\times 10^{-4}$ & ${-5.289\times 10^{-4}}_{-3.9\times 10^{-5}}^{+4.6\times 10^{-5}}$ & $-6.163\times 10^{-4}$ & $-4.456\times 10^{-4}$ \\ 
\hline 
 \end{tabular} \\ 
 }
\caption{\label{stats_results} Values of best fit, means with standard deviation, and 95\% bounds for the parameters of interest. The likelihood function $\mathcal{L}$ was maximized to $-\ln{\cal L}_\mathrm{max} =1396.2$, which corresponds with a minimum value for the $\chi^2$--function of $\chi^2_{min}=2792.4\, .$}
\end{center}
\end{table*}

\begin{figure}[ht!]
\centering
  \includegraphics[width=.82\linewidth]{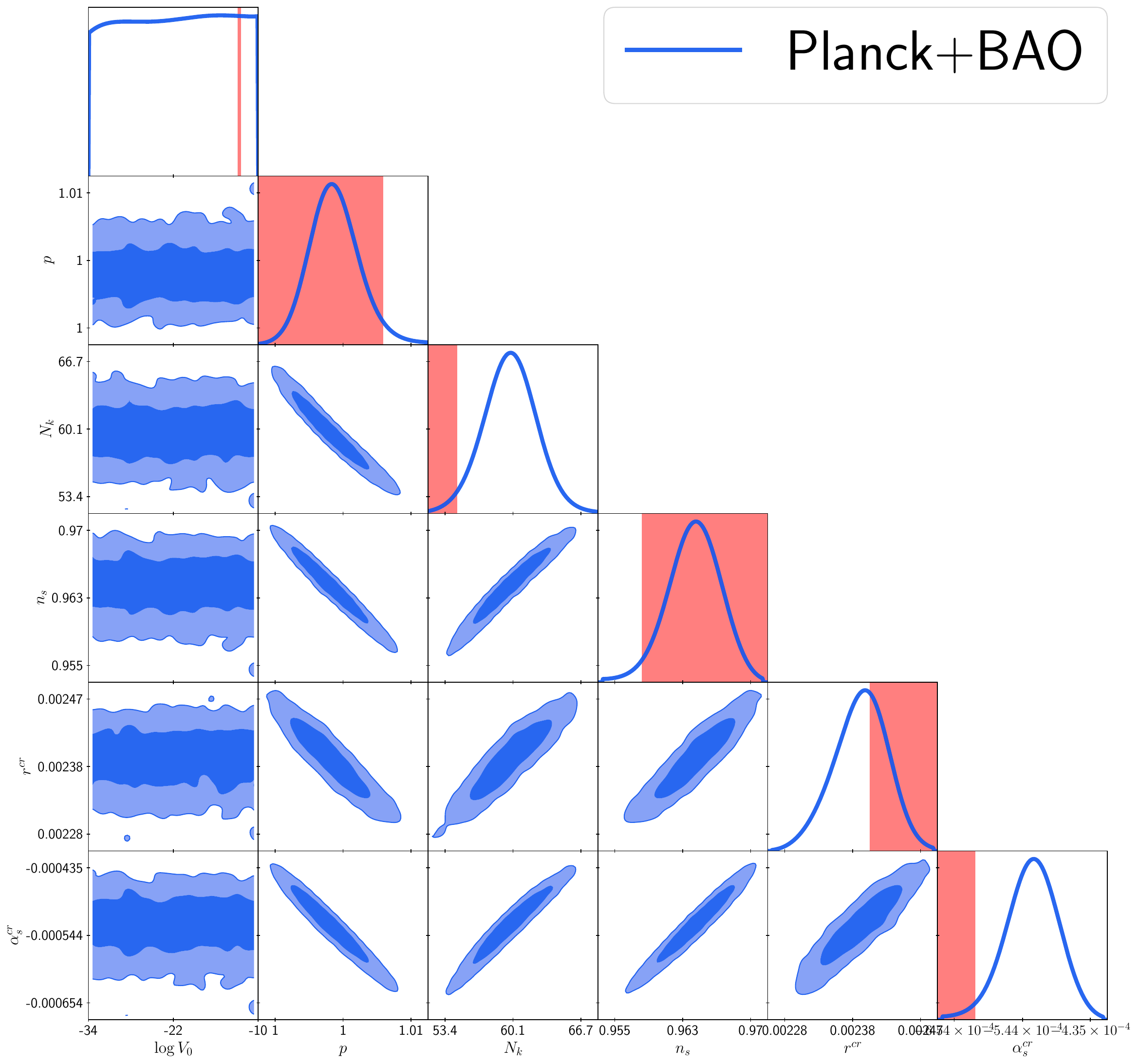}
  \caption{1D and 2D posterior distribution functions for the parameters $\log V_0\, , p\, , N_k\, , n_s\, , r^{cr}\, ,$ and $\alpha^{cr}_s$. The red vertical bars indicate the range of the theoretical values allowed from reheating given in Table~\ref{tabla}, and contained within the posteriors obtained from the statistical analysis using Planck+BAO data.}
  \label{post}
\end{figure}

In Figure~\ref{post} the 1D and 2D posteriors are shown. Besides, the theoretical bounds from reheating presented in Table~\ref{tabla} are shown in red stripes for each parameter. It can be seen that observational constraints for $\log V_0$ are broader than those imposed by reheating. On the other hand, for the potential parameter $p$, and the spectral index $n_s$ it can be observed that the theoretical bounds from reheating are thoroughly consistent with the constraints imposed by Planck+BAO. In the case of the tensor--to--scalar ratio $r^{cr}$, the lower theoretical bound of $0.0024$ is within the $1\sigma$ region. However combined data Planck+BAO indicate that lower values than that are allowed.
%%%%%%%%%%%%%%%
%%%%%%%%%%%%%%%
%%%%%%%%%%%%%%%
%%%%%%%%%%%%%%%

It can be seen that the $e$-folds number $N_k$ is constrained within a region that excludes part of the theoretical  bounds from reheating for such parameter. As already seen in Ref.~\cite{German:2022sjd}, $N_k$ depends monotonically on $\omega_{\rm re}$ such that lower values of $N_k$ are preferred for small $\omega_{\rm re}$. Similarly occurs for the running of the spectral index $\alpha_s^{cr}$, where the bounds from reheating are in a low probability region of the posterior.

\begin{figure}[ht!]
\centering
  \includegraphics[width=.82\linewidth]{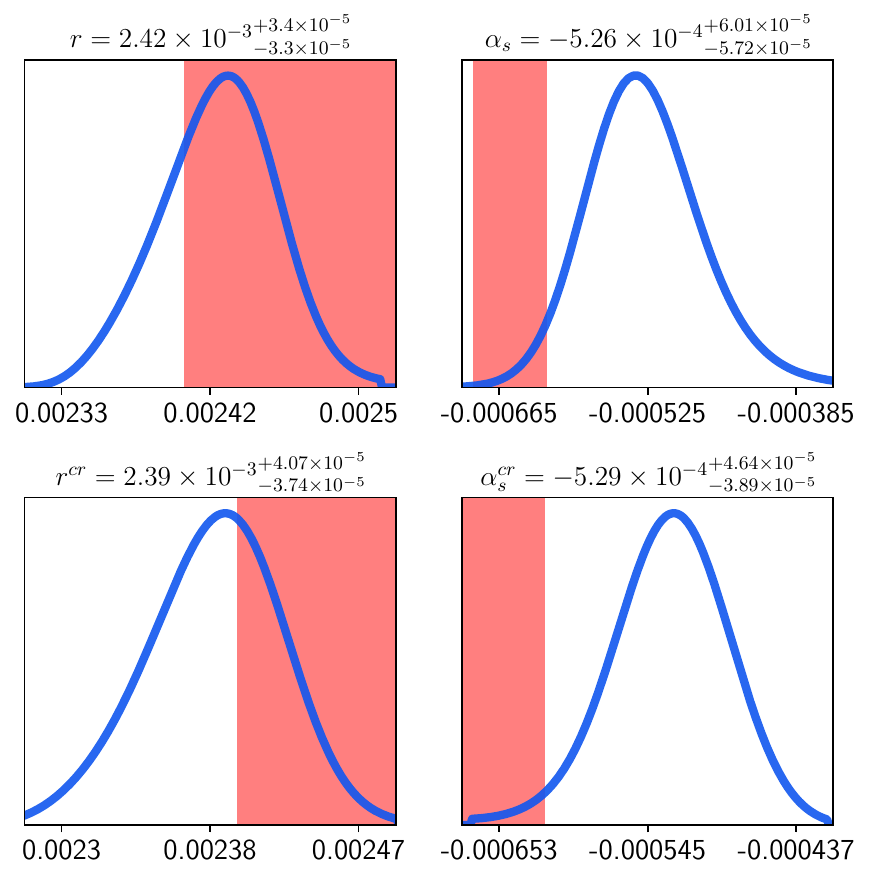}
  \caption{1D posterior distribution functions for the parameters $r\, , \alpha_s\, ,$ obtained from MCMC (top) and $r^{cr}\, ,\alpha^{cr}_s$ from the consistency relations (bottom). As in Figure~\ref{post}, the experiments constraining $r$ and $\alpha_s$ are Planck+BAO (see footnote 1), and red stripes indicate the range of the theoretical bounds from reheating given in Table~\ref{tabla}.}
  \label{post_r_alpha}
\end{figure}

In order to extend our analysis for the parameters obtained from the consistency relations, we have performed MCMC for $r$ and $\alpha_s$. The consistency relation~\eqref{genalpha} as well as the MCMC method give a negative mean value for the running. Specifically, we have obtained that the mean value for $\alpha_s$ from the MCMC is given by $\alpha_s = -5.26\times 10^{-4}\, .$ On the other hand, the tensor--to--scalar ratio as a cosmological parameter gets constrained with the value $r =2.42\times 10^{-3}\, ,$ which is in agreement with the value obtained from the consistency relation~\eqref{genr} (see Table~\ref{stats_results}). For the sake of comparison, we show in Figure~\ref{post_r_alpha} the 1D posteriors for $r$ and $\alpha_s$ obtained from the MCMC, as well as $r^{cr}$ and $\alpha_s^{cr}$ obtained from the consistency relations~\eqref{genr} and~\eqref{genalpha}.
%%%%%%%%%%%%%%%
%%%%%%%%%%%%%%%
%%%%%%%%%%%%%%%
%%%%%%%%%%%%%%%

In both cases we find a statistical agreement: the bounds imposed by reheating are contained within the posterior distributions obtained from any of the two methods (MCMC and consistency relations). In the case of the tensor--to--scalar ratio,  the result from the MCMC $r$ as well as from the consistency relation~\eqref{genr} $r^{cr}$ are statistically consistent, and both of them contain at $1\sigma$ the lower value from the theoretical bound shown in Table~\ref{tabla}. For the running $\alpha_s$ we can see that the bounds from reheating are contained within the posterior, although such bounds lie at lower probability of the posterior distribution for such parameter.

%%%%%%%%%%%%%%%%%%%%%%%%%%%%%%%%%%%%%%%%%%%%
%%%%%%%%%%%%%%%%%%%%%%%%%%%%%%%%%%%%%%%%%%%%

%%%%%%%%%%%%%%%%%%%%%%%%%%%%%%%%%%%%%%%%%%%%
%%%%%%%%%%%%%%%%%%%%%%%%%%%%%%%%%%%%%%%%%%%%

\section{Discussion and Conclusions}\label{Con}

The dynamics of the early universe is ruled by an inflationary era followed by a reheating process prior to the radiation domination. Whereas the physics of inflation has been constrained by observations such as CMB anisotropies, models of reheating are not always considered as the transitional phase at the end of inflation. In this work we have considered theoretical bounds imposed by the physics of reheating in order to analyze a generalized Starobinsky potential for primordial inflation.

We foud that the information inferred from reheating leads to more restrictive values of the cosmological parameters than those imposed by Planck+BAO data. It is then important to highlight that even when analytical and semi--analytical treatment of both inflation and reheating provide insights on the most likely values of the cosmological parameters, the implementation of statistical methods using up-to-date data is mandatory if one intends to  address the viability of inflationary potentials including physics from reheating. Is in this way in which we were able not only to infer the most likely value of the parameter $p$ generalizing the Starobinsky potential, but to quantify the level of agreement of the theoretical predictions from reheating with such observational constraints as well.

Particularly, in light of data from CMB and BAO we found tighter constraints for $p$ and $N_k$ than those presented in Ref.~\cite{Renzi:2019ewp}\footnote{It is worth mentioning that in~\cite{Renzi:2019ewp} the likelihood function for CMB data they used was Planck 2015, whereas we have used the latest data release from Planck Satellite Collaboration 2018.}. The value of the parameter generalizing the standard Starobinsky potential $p$ is statistically consistent with $p=1$, and then, the original form of this inflationary potential is preferred by the combination of data Planck+BAO. In fact, we do not perform any model comparison technique due to the small departure from the Starobinsky potential. In the case of the number of $e$-folds during inflation $N_k$, the generalized Starobinsky potential is consistent with an approximate value of $N_k\sim 60$. 
This value lies near the maximum limit imposed by Planck observations for the standard $R^{2}$ inflationary model, where the range $49 < N_k < 59$ at $95\%$ CL is reported~\cite{Planck:2018jri}.

%In the present work, broadening the range of $p$ takes us to $53<N_k<67$. However, our statistical constraints on $p$ result in a very narrow range around 1; thus, as detailed in the Appendix of \cite{Garcia:2023tkk},   the value  $T_{re}=3.1\times 10^9\,\mathrm{GeV}$ \cite{Gorbunov:2010bn} is still adequate.

Concerning inflationary observational parameters such as the scalar spectral index $n_s$ and the scalar amplitude of perturbations $A_s$ we obtain the following: our inferred value for $n_s$ (its mean value and standard deviation) lies within the $1\sigma$ CL reported by~\cite{Planck:2018vyg} ($n_s^{Planck} = 0.965\pm 0.004$). On the other hand, since $A_s$ is proportional to the amplitude of the inflaton potential $V_0$ (see Eq.~\eqref{IA}), we note that the values obtained from the posterior of $\log V_0$ are consistent with $10^9A_s = 2.1^{+0.0107}_{-0.0112}$, which is also in agreement with the value reported by Planck 2018 of $A_s^{Planck} = 2.105 \pm 0.030$ (see last column of Table 2 from~\cite{Planck:2018vyg} where the $68\%$ limits is shown considering the same Planck+BAO set of data we have used in the present work).

The use of consistency relations is useful to reduce the number of degrees of freedom in the parameter space. The model--dependent nature of these relations allows us to test the generalized Starobinsky potential in a self--contained way, and to put constraints on parameters such as $r^{cr}$ and $\alpha_s^{cr}$ at low computational cost. Even when the considerations from reheating have been tested indirectly via theoretical constraints from the semi--analytical approach, it was possible to quantify and compare the results coming from such epoch with CMB observations, which constitutes the most robust and precise probe of the early universe. Both inflation and reheating leave characteristic signatures in the spectrum of primordial gravitational waves that could be detected in the future~\cite{Kawamura:2006up,LiteBIRD:2022cnt}, and hence the importance in constraining theoretical scenarios in order to learn more about the physics ruling the dynamics of the earliest times of our universe. 

%%%%%%%%%%%%%%%%%%%%%%%%%%%%%%%%%%%%%%%%%%%%
%%%%%%%%%%%%%%%%%%%%%%%%%%%%%%%%%%%%%%%%%%%%

%%%%%%%%%%%%%%%%%%%%%%%%%%%%%%%%%%%%%%%%%%%%
%%%%%%%%%%%%%%%%%%%%%%%%%%%%%%%%%%%%%%%%%%%%
\acknowledgments

F.X.L.C. acknowledges Beca CONACHyT. J. C. H. acknowledges support from program UNAM-PAPIIT, grant IG102123 “Laboratorio de Modelos y Datos (LAMOD) para proyectos de Investigación Científica: Censos Astrofísicos". as well as the sponsorship from CONAHCyT Network Project No. 304001 “Estudio de campos escalares con aplicaciones en cosmología y astrofísica”, and through grant CB-2016-282569.
%%%%%%%%%%%%%%%%%%%%%%%%%%%%%%%%%%%%%%%%%%%%
%%%%%%%%%%%%%%%%%%%%%%%%%%%%%%%%%%%%%%%%%%%%

\bibliographystyle{plunsrt}
\bibliography{bib}

\end{document}